\documentclass{article}
\usepackage[utf8]{inputenc}
\usepackage{graphicx}
\usepackage{lineno}
\usepackage{soul}
\usepackage[margin=1.0in]{geometry}
\usepackage{float} 
\usepackage{xcolor}
\usepackage{tocloft}

\usepackage[backend=biber,style=vancouver]{biblatex}
\title{Wearable Tracking of Eye and Body Movements During Breaching Training: Towards Real-Time Blast Injury Monitoring }
\author{Jeremy Kemmerer$^1$, James R. Williamson$^1$, Joseph Kim$^1$, Elizabeth Halford$^2,$ \\  Hrishikesh M. Rao$^1$, Christopher J. Smalt$^1$\thanks{Corresponding Author: \emph{Christopher.Smalt@LL.mit.edu}}}

\date{%
    \small{$^1$Human Health \& Performance Systems Group, MIT Lincoln Laboratory, Lexington, MA, USA\\
    $^2$Cardea Project Management, LLC., Alexandria, VA, USA\\
     }}

\begin{document}

\maketitle

\begin{abstract}
% One major challenge that remains is to define occupational exposure limits given the uncertainty around the units / components of the explosion that contribute to injury. Likewise, the corresponding safe exposure levels are not known and may vary between individuals. In this work, we describe a system to simultaneously monitor physiology and blast exposure levels and demonstrate how this system can identify individualized exposure levels corresponding to acute physiological response to blast exposure.

Repeated exposure to blast overpressure in occupational settings has been associated with changes in cognitive and psychological health, as well as deficits in neurosensory subsystems. In this work, we describe a wearable system to simultaneously monitor physiology and blast exposure levels and demonstrate how this system can identify individualized exposure levels corresponding to acute physiological response to blast exposure. Machine learning was used to develop a dose-response model that fused multiple physiological measures (electrooculuography, gait, and balance) into a single risk score by predicting the level of blast exposure on held-out subjects (Fused model, $R=0.60$). We found that blast events with peak pressure levels as low as 0.25 psi could be related to physiological changes and hence may contribute to blast injury. We also identified an individual subject with deteriorating reaction time scores that consistently showed a rapid and anomalous change in physiology-based risk scores after exposure to low-level blast events. Our results suggest that the wearable approach to blast monitoring is viable in weapons training environments as a complement to more direct but sparsely administered brain health assessments, potentially viable in austere environments, and that fusing multiple physiological signals can improve sensitivity.  

\end{abstract}

\section{Introduction}

Many military and law enforcement personnel are routinely exposed to repeated blast overpressure in training and operational settings.  Often, these exposures are referred to as low-level blast (LLB) and can be the result of firing one's own weapon system, and are considered subconcussive, i.e. not causing a loss of consciousness.  The range of what constitutes LLB in pounds per square inch (psi) is not well defined, but a range of 3-6 psi or less is often used in the literature \cite{woodall2023repetitive,carr2016repeated}.  The relationship between LLB and negative brain health outcomes is also unclear, but several studies have reported significant changes in military populations \cite{phipps2020characteristics, carr2021blast}.  Neurosensory systems in the brain in particular, such as oculomotor \cite{woodall2023repetitive,capo2015effects,rao2023changes}, vestibular \cite{akin2017vestibular,williamson2022using}, and auditory function \cite{kulinski2023acute,hecht2019characterization}, are believed to be particularly sensitive to LLB.  Overexposed individuals are also believed to have an increased risk of developing anxiety disorders, depression, and mental health disorders \cite{walker2015prevalence,belding2023single} and headaches \cite{sajja2019role}. 
%A review paper to cite: \textbf{//CHECK//}

Despite evidence of brain health risks of blast exposure, it remains challenging to acutely detect or prevent injury and establish exposure limit guidelines.   Currently, the US military guideline for maximum exposure sets a limit of 4 psi \cite{Hicks2024BlastMemo} for blast overpressure peak pressure. 
Historically, this limit was derived from auditory injury, specifically the risk of rupture of the tympanic membrane \cite{ritenour2008tympanic,hickman2018blast}.  In providing recent exposure limit guidance, the US Deputy Secretary of Defense also acknowledges that the underlying medical science is evolving and that future updates will be needed \cite{Hicks2024BlastMemo}.  One major limitation of a maximum peak pressure threshold is that it does not specify the number of rounds or exposures that are acceptable; according to this standard, an individual can acceptably be exposed to an unlimited number of rounds at 3.99 psi, but a single exposure at 4.00 psi is considered unacceptable \cite{smith2024end}.   Although an important first step, a single exposure threshold is incomplete as a safety measure to protect brain health in environments where repeated lower-level exposures occur.  

Personal blast exposure has to-date been quantified in multiple ways.  An individual's history of blast exposure can be estimated based on questionnaires or service history (e.g., GBEV) \cite{modica2021development}. Additionally, there are three main approaches for more directly quantifying short-term blast exposure: estimating or counting the number and type of rounds fired for a specific weapon, measuring blast pressures directly at fixed locations, or measuring blast pressures on the body. In other occupational sound exposure use cases, such as industrial noise exposure, a dosimetry-based approach is often preferred, where each individual has a device to measure the exposure \cite{goodwin1998occupational, smalt2017noise}.  Personal exposure dosimetry can also account for the complexity of the environments where blast exposure occurs, including factors such as weapon type, physical proximity to the blast event, or effects of rigid structures such as walls or enclosures, which can significantly impact blast overpressure levels \cite{davis2019ear, smalt2017noise}. Personal, wearable dosimetry also enables the development of accumulating dose measures.  
% relying on a weapon-specific characterization (i.e., estimating the exposure based on the number and type of rounds fired) or measuring blast pressures directly at fixed locations or on-body.
 
Unlike a maximum single-exposure measure (e.g., peak pressure), accumulating blast exposure measures consider the time-history of exposure and incorporate both overpressure levels and the quantity of blast events. Furthermore, they can be reported as a single, cumulative score for an individual training session. Despite these apparent advantages and the growing availability of dosimetry devices that record and measure individual blast events, several challenges have limited the development and adoption of cumulative exposure measures for LLB. A primary challenge is the lack of a gold-standard brain health outcome "response" variable. Other challenges include inter-device variation in overpressure measurements, which can be significant depending on the characteristics of the sensor (i.e., sampling rate, peak pressure limits, or the placement on the body), and uncertainty regarding a threshold for significant blast events with respect to brain injury. 

Determining safe exposure limits based on cumulative dose measures requires tools that are sensitive and specific to mild brain injury or functional change as it occurs \cite{elder2014effects}.  An effective decision tool must be able to rapidly assess change in an individual's state relative to their baseline following a potentially injurious exposure. This is critical for preventing impairment and enabling recovery \cite{schmid2021review,mccrory2011sports}. Allowing individuals to continue to be exposed after injury can affect their performance in critical operational scenarios through decrements in reaction speed and executive functioning \cite{lavalle2019,carr2016repeated,carr2017perspectives} and may increase the risk of further injury.  

Recent advancements in wearable technologies and machine learning techniques provide clinicians and engineers with the opportunity to collaborate, harnessing multi-parameter sensing and signal processing for the real-time, quantitative diagnosis of mild traumatic brain injury (mTBI) during exposure.
%Recent advances in wearable technologies and machine learning methodologies offer clinicians and engineers the opportunity to work together in order to leverage multi-parameter sensing and signal processing for quantitative, real-time mTBI diagnosis that can occur during exposure.   
Wearable technologies are becoming ubiquitous and can be used to detect conditions such as fatigue \cite{adao2021fatigue} and cognitive changes or brain health disorders such as Parkinson's disease \cite{williamson2021detecting}, dementia \cite{mc2020differentiating}, and mental health \cite{kang2022wearable}. In our previous work, we used wearable gait, balance, and eye tracking via electrooculography (EOG) to detect changes due to blast exposure \cite{williamson2022using,rao2023changes}.  These approaches typically rely on machine learning techniques, which are needed to derive `physiological biomarkers' from continuous wearable sensor data \cite{schmid2021review}. The current use of wearable technology focuses mainly on blast exposure monitoring through the use of pressure sensors \cite{wiri2023dynamic}, rather than physiological status.

\begin{figure}[h!]
\centering
\includegraphics[width=130mm]{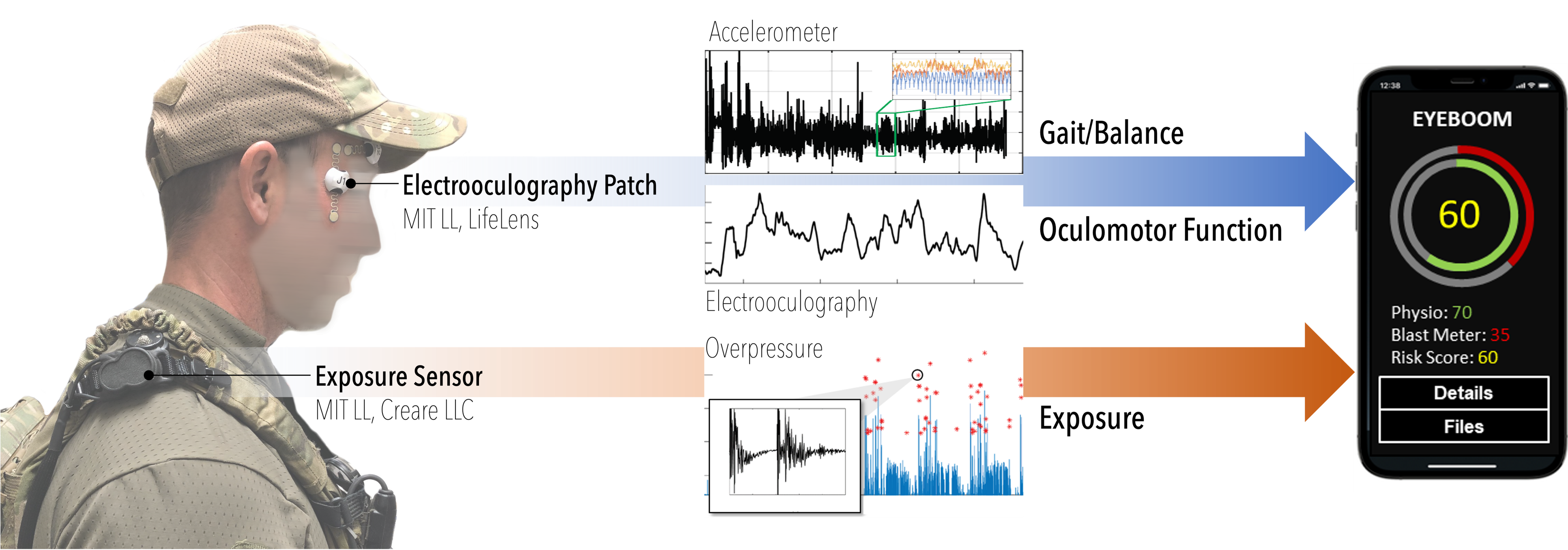}
\caption{A wearable physiology and exposure system for measuring blast exposure risk, and notional smartphone application to provide real-time, continuous, risk assessment.  A low-level-blast exposure sensor is placed on the shoulder to capture incident pressure, and an EOG patch is worn near the eye to capture eye movements and blinks, as well as gait and balance.}
\label{fig:overview}
\end{figure}

Despite the challenges and considerations involved in quantifying low-level blast exposure and brain health outcomes, a combined approach using wearables to monitor exposure and physiology has significant potential benefits.  Wearable monitoring approaches are now available that can measure physiological and cognitive status in addition to exposure, thereby providing two major benefits  \cite{schmid2021review}.  The main practical benefit of a combined physiology/exposure wearable system is that it can be used to notify an individual in real-time if their risk or physiological change exceeds a threshold even before the end of a training session, minimizing potential injury, as illustrated in Figure \ref{fig:overview}.  A second benefit is that combined dose-response data is acquired in the same system on the same person, thus providing the opportunity for iterative improvement in risk prediction models.

% To reduce the risk of potential injury, it is highly desirable to develop a dose-response risk relationship based on acute changes in cognitive function or physiological status that could be applied at the individual level. 
% As a consequence of our incomplete set of tools for acute assessment of brain injury, well-defined 

In a previous effort with the U.S. Special Forces during urban breaching training, we showed that continuous eye and body movements are individually predictive of cumulative blast exposure pressure energy.  Here, we replicate that result on a larger dataset with a similar population but 
with improved sensor systems and wearable technology suitable for austere environments.   We also develop new combined dose/response models that fuse multiple physiological measures into a single ``risk score" and correlate these scores with cumulative LLB metrics to identify thresholds for significant blast events.  Finally, we compare our results against other independent neurocognitive assessments (e.g., ANAM) to identify individuals who may be at elevated risk of injury.

% probably discussion:  . Our approach derives risk through acute dose/response models combining both cumulative exposure energy and changes in physiology, as opposed to a single peak overpressure limit.  The system could flag individual users who have both a high cumulative exposure energy AND a physiological change as being at greatest risk.
% By directly addressing the issues of individual susceptibility, notification before injury occurrence, and interpretation of blast metrics for different weapon systems, this system would provide a major step forward from previous technologies.
% It is our view that on-body, objective acoustic measurements are best suited to account for these factors, but the question remains of what the best measures are for predicting brain injury, and at what levels and number of exposures injury is likely to occur.

\section{Methods}

\subsection{Study Population}

Special Forces training environments offer a unique opportunity to monitor the effects of repetitive exposure to overpressure to prevent brain injury. Instructors or "cadre", who have prior experience in close-quarters battle (CQB) techniques including interior and exterior breaching, are exposed to a significant number of low-level blast overpressure events while observing trainees. They can serve on an assignment spanning multiple years and are responsible for instructing multiple courses per year. In addition to overpressure exposure from explosives, instructors are in close proximity to small-arms fire and other munitions. A total of 28 instructors participated in a voluntary data collection effort as part of a program evaluation effort, which included overpressure dosimetry, oculomotor function, hearing function, symptom reports, and cognitive performance monitoring. In this work, we report the results of a secondary analysis of a subset of these de-identified datasets.

\subsection{Blast Overpressure Monitoring}

Special forces instructors donned a low-level blast monitoring device (MNOISE) to continuously monitor exposure to low-level blast overpressure \cite{smalt2022development}. This device continuously monitors time-average sound pressure levels (e.g., LZeq) using a custom two-microphone setup to increase the dynamic range, with a maximum peak pressure limit of 5.2 PSI (185 dBP). Overpressure events above 0.03 PSI / 140 dB SPL trigger the recording of two raw waveforms, one from each microphone that are recorded in units of Pascals. Both continuous sound levels and blast overpressure event recordings are extracted as files from the MNOISE and processed using custom MATLAB code to compute cumulative blast exposure measures \cite{smalt2022development}. 

Artifacts due to physical knocking or electrical artifact can potentially be a significant source of contamination in blast dosimetery data. In this study, we detected and removed artifacts from our dataset as a pre-processing step. Our approach used a classifier to check for consistency between the two microphones. Whereas real sound events have a close correspondence between microphones in high- and low-pressure recordings, artifacts are generally random and uncorrelated between microphone channels. Sessions with too many detected artifacts (e.g., greater than 20) were discarded due to the potential for remaining undetected artifacts to be confused as significant blast events (15 percent of sessions were removed this way).

Cumulative measures of exposure were computed from multiple typical per-event acoustic measures (i.e., LZeq8hr, Peak Impulse, Peak Pressure, and supra-threshold event index) to generate accumulating measures and cumulative session totals (i.e., LZeq8hr, Cumulative Peak Impulse, Cumulative Peak Pressure, Cumulative Blast Count). Both accumulating and total measures for each session were compared to physiology.

\subsection{Continuous Movement and Eye Tracking}

Alongside the MNOISE, instructors donned a LifeLens electrode patch and sensor hub, together referred to as a "kit", (Figure \ref{fig:overview}). The custom LifeLens patch adheres directly to the skin and can be used for several hours continuously. A sensor hub attaches to the patch that contains an accelerometer and records electrical signals from the body. Here, we placed the patch and hub next to the eye in vertical and/or horizontal orientation for eye tracking via EOG measurements. Subjects were given the option to wear one or both patches on each eye, resulting a few distinct datasets: two vertical EOG measurements for each eye, one vertical and one horizontal measurement for each eye, and in a few cases, two horizontal patches, one above each eye. Depending on the choice of orientation, blink features (vertical orientation) and saccade features (both vertical and horizontal) were extracted from the continuous EOG signals.
LifeLens kits were generally applied 30 minutes to 1 hour before the start of each day's training session, and were removed at the end of the session. In some cases, individuals wore the kits overnight and throughout the course of training the next day. Accelerometry and eye-tracking data were later extracted after the sensor was removed and stored for post-processing using a custom LifeLens data interface. 

\subsection{ANAM Assessments}

The Automated Neuropsychological Assessment Metrics (ANAM) is a computerized cognitive test battery that is used to measure the cognitive effects of injury, exposure, or illness \cite{ANAMManual}. The ANAM consists of a flexible library of cognitive tests and behavioral questionnaires that provide performance-based measures of neuropsychological functions. A version of ANAM is approved for widespread use in DoD, is administered on a laptop, and takes approximately 25-30 minutes to complete. A condensed, seven-minute battery was offered to the subjects in this study on a tablet computer that was generally administered both before and after training. The tests included Simple Reaction Time, Go/No-Go, Spatial Processing, and Memory Search. This ANAM battery was voluntary and self-administered.

\subsection{Gait and Balance Features}

Gait and balance features were computed from 100 Hz 3-axis accelerometry data from the vertically oriented LifeLens sensor, using methods similar to those described in \cite{williamson2022using}. 
Accelerometry data were segmented into gait and low-movement (LM) periods to distinguish changes over time in head movements that occur during ambulation from those that occur during periods of standing or sitting.

Gait and LM bouts were segmented primarily based on the variance of acceleration magnitudes, computed in sliding 10 s windows. 
Gait bouts were detected when windowed variances are suprathreshold ($\sigma^2_m(t) > 0.03$). Suprathreshold time intervals were then joined together if separated by subthreshold intervals of 15~s or less. The resulting gait bouts were then divided into contiguous 5~s frames. Gait features were extracted from a gait frame only if it passed a periodicity test: The first principal component of the three accelerometry axes in the 5~s window was computed, and its autocorrelation peak within a plausible step duration range of 0.35 s to 0.85 s was required to have a height and peak prominence of at least 0.2.

The gait features extracted from each 5 s frame were based on the eigenspectra of high-dimensional correlation matrices that were constructed with time delay embedding (TDE) at five delay scales. There are 7 time delays at each scale, which when combined with 3 accelerometry axes results in correlation matrices with dimensionality $21 \times 21$. The 5 delay scales contain delay spacings of 30, 70, 150, 310, and 610 ms.
In a given delay scale, the TDE eigenspectra represent acceleration time-frequency dynamics via the shape of the time-delay scatter at that scale.
By using multiple scales, the method represents dynamics at multiple time-frequency bands. The TDE technique is described in \cite{williamson2022using}. 
The TDE features, concatenated across 5 delay scales, have total dimensionality of 105. This dimensionality was reduced using PCA into 20 principal components, which explain $99\%$ percent of the total variance in TDE features.

LM bouts were detected when windowed variances fell between two thresholds, $0.001 < \sigma_m(t)^2 < 0.01$, across an interval of at least 30 s. These thresholds constrain activity to a lower level of acceleration variance, which is consistent with low-movement activities such as standing or sitting. Balance dynamics were quantified using the acceleration path length feature, which was extracted from contiguous 5 s frame within LM bouts. Given a set of $n=500$ accelerometry vectors in a single frame, $\{x_i,y_i,z_i\}$, the path length (PL) for that frame is \begin{eqnarray}
PL = \sum_{i=2}^n \sqrt{ (x_i-x_{i-1})^2 + (y_i-y_{i-1})^2 + (z_i-z_{i-1})^2 }.
\end{eqnarray}
Because LM bouts were segmented based on a fixed range of acceleration magnitude variance, higher PL values should occur when a greater proportion of dynamics occurs at high time frequencies.

\subsection{Eye Tracking Features}

Blinks and saccades were detected from the 500 Hz EOG signals. Blinks were detected using a single vertically oriented EOG, and saccades from paired vertical and horizontal EOG. First, artifacts were removed from EOG signals as follows. The signals were bandpass filtered from 0.1 to 10 Hz using a third-order Butterworth filter. This frequency range was selected to enable detection of blinks, which have a typical duration of 100-400 ms, while attenuating EOG activity at higher frequencies \cite{nystrom2024blink}.
Periods of the filtered signal where the power was greater than 2.5 times the 5-minute moving median were marked as artifacts and excluded from analysis, a method derived from \cite{brunner1996muscle}.

Next, blinks durations were extracted as follows. The bandpassed vertical EOG was processed using a continuous wavelet transform with a Haar wavelet at scale 80, a resolution suited for the frequency range of blink signals \cite{naga2012denoising}. Negative and positive peaks of the wavelet transform were detected based on a minimum peak width of 200 ms and a height (in negative and positive directions) above the 95th percentile. Matched negative and positive peaks were identified as blinks if they were separated by less than 200 ms. These parameters were selected to maximize the ratio of true blink detections to false positives in an internal analysis of a subset of hand-labeled data. For identified blinks, the feature used by the prediction algorithm was the blink duration, defined as the duration of the local EOG peak at half the peak amplitude, based on bandpass filtering of the original EOG signal between 0.1 and 25 Hz. 

Saccade features were obtained from a subset of subjects who wore paired vertical and horizontal EOG sensors. The same bandpass filtering and artifact removal was done as with blink processing. The horizontal filtered EOG signal was linearly interpolated to be synchronized with the vertical EOG signal.
First, the general time points of potential saccades were detected where change points in the horizontal filtered EOG were found, based on a threshold equal to the global standard deviation of the signal. Next, the horizontal signal was smoothed using a 50 ms moving average, and the absolute velocity was computed based on the absolute value of the discrete first-order derivative. The precise saccade time points were determined based on the maximum absolute horizontal velocity within a 50 ms window.
The saccade amplitude was computed as the absolute difference between the filtered horizontal EOG at the start and stop of the saccade window. If the saccade amplitude was less than 0.25 times the global standard deviation of that signal, and if the correlation between the filtered horizontal and vertical EOG signals within the saccade interval was greater than 0.6, then the saccade was positively identified. The saccade amplitude feature was used by the prediction algorithm. 

%EOG datasets were cleaned (low-pass filter) and blinks events were extracted using wavelet processing. Multiple measures or features were computed (e.g. duration and rate) from detected blinks  [ref Military Med paper]. For sessions with cadre who wore a second electrode and sensor hub in a horizontal orientation, saccade features were extracted from each combined vertical and horizontal EOG dataset.

\subsection{Feature Change Score}
\label{sec:feature-change}

A method is next implemented to register changes in gait, balance, blink, and saccade features over time, relative to the recent history of variability for each feature.
First, the features were smoothed to capture a running estimate of their local-time mean value. Gait and balance features were extracted opportunistically from temporally dispersed gait and LM frames. The running mean of these features was computed as the average across the most recent 30 gait or LM frames.
Eye blink and saccade features were computed more frequently, based on the detection of eye blinks and saccades. The running mean of these features was computed across the most recent 20 blinks or saccades.

Next, individualized changes in the features were computed over time, relative to the mean and standard deviation of the features from each subject's recent history. This on-line change-detection technique was originally developed for detecting physiological changes that predict heat stroke \cite{buller2021real}.
In this method, on-line z-scoring is computed to detect change in a time series variable, $f(n)$, where $n$ indexes the time step number. Change in the variable relative to it's mean and standard deviation is quantified by using recursive filtering to track the first and second moments of the time series, $f_1(n)$ and $f_2(n)$. Then, $f(n)$ is mapped into an instantaneous change score via
\begin{eqnarray}
f_z(n) = \frac{f(n)-f_1(n)}{\sqrt{f_2(n)-f_1(n)^2+\frac{1}{w(n)}}},
\end{eqnarray}
where the count variable, $w(n)$, is included in the denominator to prevent large fluctuations when the number of data points is small.
Finally, the instantaneous feature change score is smoothed to reduce sensitivity to short-term fluctuations, producing the feature change score that provides input to the regression model. Smoothing is done using recursive filtering, with $\alpha = 0.0001$, 
\begin{eqnarray}
f_{zs}(n) = \alpha f_z(n) + (1-\alpha) f_{zs}(n-1),
\end{eqnarray}

\subsection{Mapping Feature Changes into Risk Scores}
\label{sec:regression-model}

The ability to predict blast exposure level from feature change scores was evaluated using leave-one-subject-out cross-validation training and testing of a regression model. All sessions from the same subject were assigned exclusively either to a training or a testing fold.
A session was included for each modality only if the feature data spanned at least a one hour duration. In addition, gait and balance sessions were included only if there were at least 25 gait and balance frames detected.
Saccade features were extracted only in the subset (roughly half) of sessions in which there were two orthogonally positioned EOG sensors (horizontal above the eye and vertical near the temple).

The regression model uses an ensemble of Gaussian mixture models (GMMs), in which each element of the ensemble is trained to discriminate {\em higher} exposure from {\em lower} exposure, based on multiple  exposure-level thresholds that partition the training data into two classes. The ensemble across the set of training data partitions is a GMM staircase \cite{williamson2022using}. The same GMM parameters were used as in \cite{williamson2022using}, except each GMM was defined with 5 Gaussian components instead of 10. 

Seven GMM staircase partitions were used based on the following percentile thresholds of the blast exposure outcome measures, $y_t$, in the training set: 
$\{ 12.5, 25, 37.5, 50, 62.5, 75, 87.5\}$. The risk score produced by the GMM staircase for each feature modality is a log-likelihood ratio, which is based on the sum of likelihoods across the ensemble of {\em higher} and {\em lower} exposure models, trained on the set of $y_t$ thresholds.

\subsection{Multimodal Fusion}
\begin{figure}[h]
    \centering
    \includegraphics[width=0.95\linewidth]{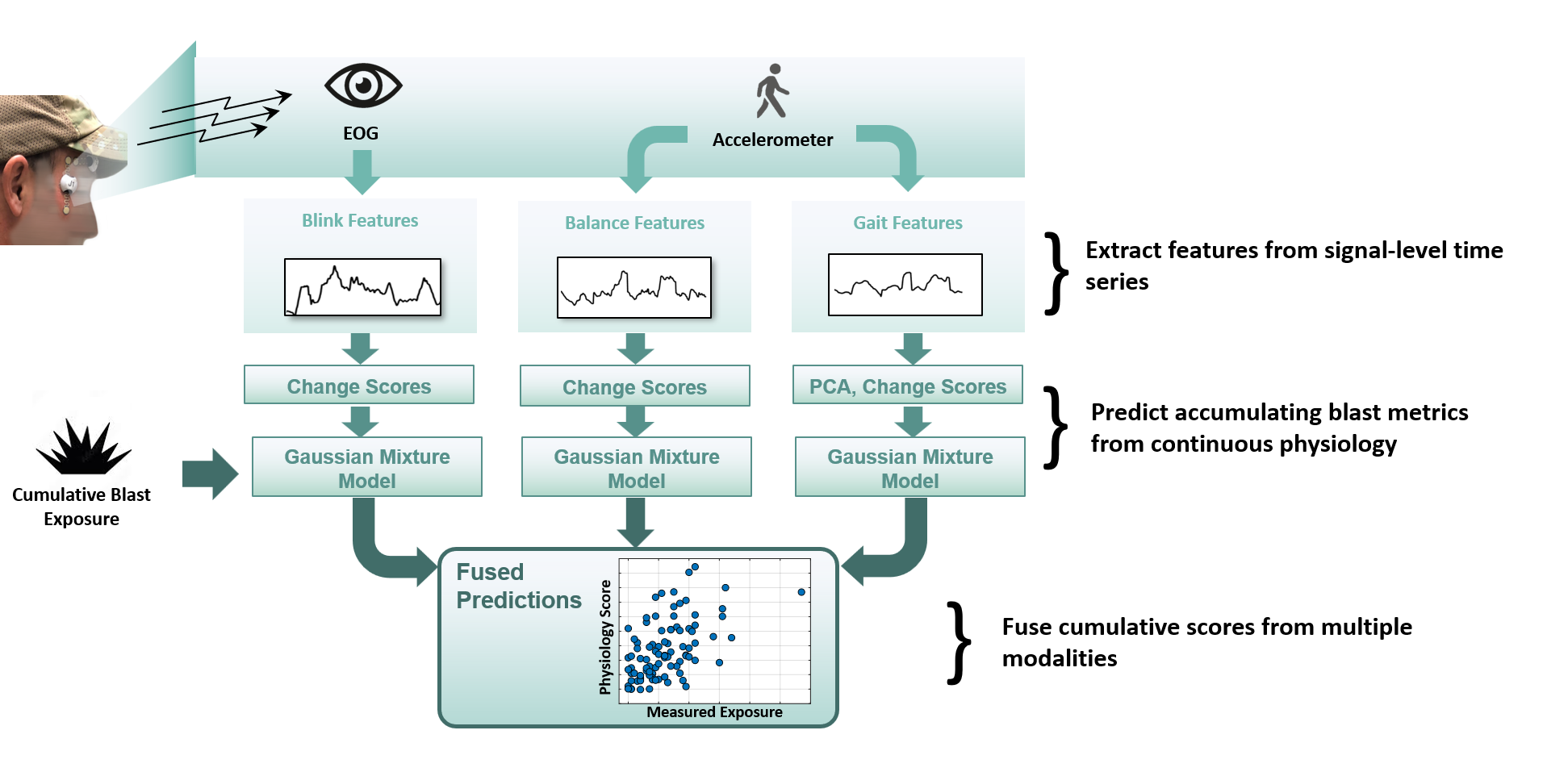}
    \caption{Multi-modal feature extraction, model training, and fused scores. Physiology time series are extracted for each modality and used to predict exposure (score) time-series using an individualized model. Per-session scores are fused to compute dose/response correlation. }
    \label{fig:training_fusion_diagram}
\end{figure}

Continuous risk scores were computed for each individual feature (e.g., blink rate) and subsequently fused at the session level (Figure \ref{fig:training_fusion_diagram}). Fusion was done by averaging the risk scores (log-likelihood ratios) from all features available for each session. 
There were 91 sessions selected for the fusion analysis, based on sessions for which there was sufficient data for at least two feature modalities. Of these, there were 91 sessions with both balance and blink duration features, 79 sessions with gait features, and 36 sessions with saccade features.
\clearpage

\section{Results}
\subsection{Comparison of Physiological Response and Dose with Reaction Time Scores} 
To determine post-hoc which of the cadre may have experienced brain injury, we first performed a linear regression on the time-series of ANAM reaction-time scores for each subject over a roughly 3-year period spanning the sensor data collection. We identified 4 of 29 cadre with ANAM test scores having a statistically significant (nonzero, p $<$ 1e-3) and positive trend of reaction times, indicating long-term deterioration. Three of these four subjects wore dosimeters and LifeLens sensor kits, making it possible to compare ANAM and continuous monitoring data. 

\begin{figure}[h]
    \includegraphics[width=\linewidth]{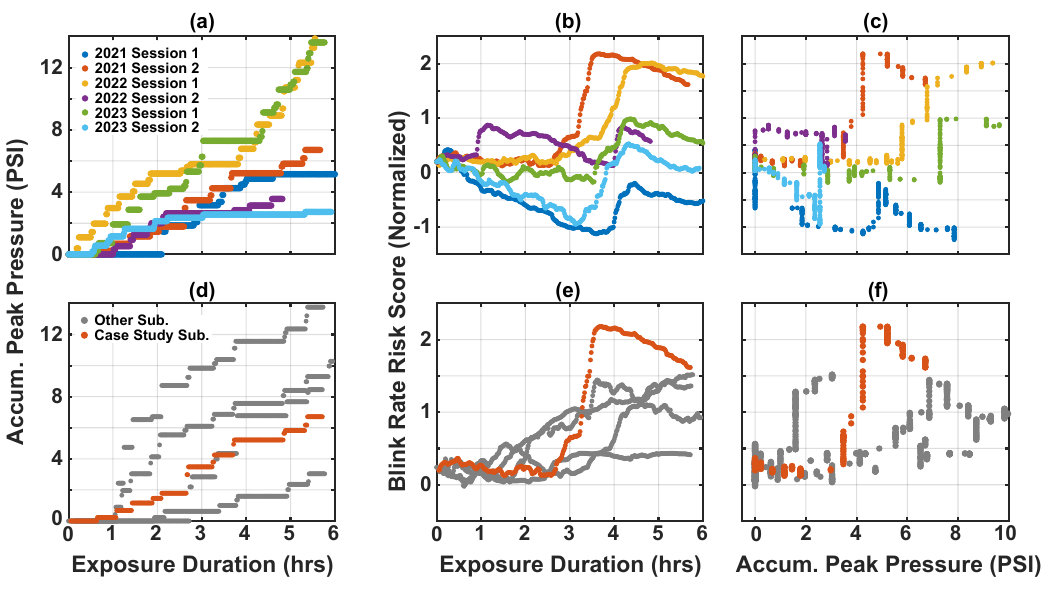}
    \caption{Case study participant blink scores versus time and exposure across several training days (bottom row) and exposure and blink scores for multiple subjects on a single training day (top row). Each plot color in the top row corresponds to a unique participant. The case study subject (indicated by in order in the bottom row) showed a rapid increase in blink scores with respect to time in (e) but a relatively mild exposure in (d). Similar rapid increases in physiology for this subject can be seen on other training days that span multiple years in (b) and appear as near-vertical lines versus exposure (d). The session indicated in orange in the top row corresponding to the same training day as the data in the bottom row.} 
    \label{fig:Case_Study}
\end{figure}

Next, for each of these three subjects, we examined their daily physiology score time series, computed as described in the previous section. For one of these subjects, we found a rapid 
%(relative to other cadre) 
increase in blink and balance scores during most of their training sessions. We note that the other two subjects with ANAM changes did not show a rapid change in their physiological data. Also,  two of the case study subjects sessions did not show a rapid increase in scores, but in both cases their sessions were truncated in duration (e.g., ended after an exposure duration of less than 3 hours, compared to a 6-hour typical duration).
Figure~\ref{fig:Case_Study}  shows this subject's blink scores from multiple daily sessions, as a function of time (b) and of exposure level (c). Exposure level is plotted as an accumulating time series in (a).
One of these daily exposures and blink scores is shown (orange) along with other subjects' exposure and physiology for a single training day in Figure~\ref{fig:Case_Study} (d-f). 

 On average, rapid increases occurred at a cumulative peak pressure of about 4.5 PSI or a cumulative exposure count of 12 low-level blast exposures (threshold peak pressure of greater than 155 dB SPL). While less apparent, a similar trend was noted in the balance scores. Operationally, these findings were of concern to the range commander, who already previously identified this subject as having significant history of previous exposure. None of the remaining cadre in the cohort consistently showed a similar trend. As can be see in Figure~\ref{fig:Case_Study} (d), this subject's measured exposure was unremarkable: at or slightly below average compared to four other subjects on the same day.

%at a cumulative blast count of 12.8 blast events or a cumulative peak pressure of 4.48 PSI (both measures with a 155 dB SPL blast event threshold). 

\subsection{Dose-Response Models for Significant Blast Events}
\label{sec:dose-response}

The recorded overpressure waveforms captured by the MNOISE dosimeters were processed as previously described to generate four candidate accumulating dose measures -- LZeq8hr, Peak Impulse, Peak Pressure, and Blast Count -- and each at seven peak pressure threshold levels -- 140, 145, 150, 155, 160, 165, 170 dB SPL. Recorded events with a peak pressure below each corresponding threshold level were excluded from the metric. As a result, measures with higher thresholds included fewer (but more significant) blast events. 

The cumulative totals of the four dose measures were compared in pairs to assess their similarity. Table \ref{tab:blast_metrics_spearman}  compares cumulative blast measures with a maximum pressure threshold of 160 dB SPL (0.25 PSI). Two pairs of measures were found to be highly correlated (Spearman's correlation $>$ 0.9): level-based measures (accumulating the count of suprathreshold blast events and accumulating the total peak pressure of the events), and energy-based measures (accumulating positive impulse and time-varying LZeq8hr). Combinations of level-based and energy-based measures showed lower correlations. Taking into account the impact of the threshold level, the correlation coefficients for the same measures at different thresholds were as low as 0.4. These results suggest the choice of accumulating exposure measure and threshold level could significantly impact model predictions, and therefore a complete set of measures and thresholds was analyzed.  

\begin{figure}
    \centering
    \includegraphics[width=1\linewidth]{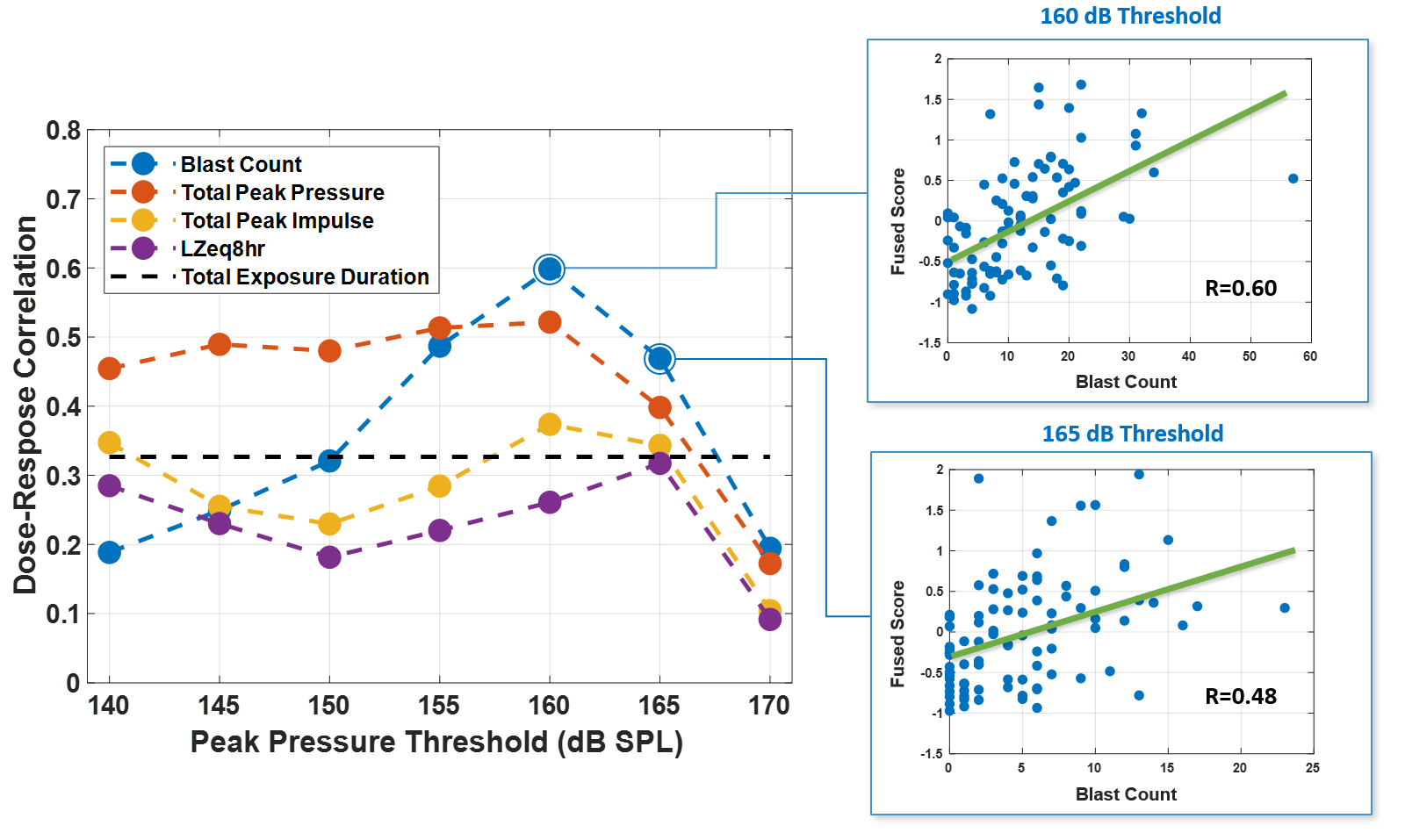}
    \caption{Dose-response correlation versus minimum peak pressure threshold for four cumulative overpressure measures. A threshold of 160 dB SPL (0.25 PSI) showed the highest correlations with physiology-based risk scores, suggesting events at this level (and above) are important to predict the physiological effects of blast exposure.}
    \label{fig:DoseResponseThreshold}
\end{figure}

%\begin{figure}
%    \centering
%    \includegraphics[width=0.75\linewidth]{MetricCorrelations_160dBSPL.png}
%    \caption{Scatter plots of per-session cumulative blast metrics. Level-based and energy-based measures were found to have high correlation coefficients}
%    \label{fig:blast_metrics_scatter}
%\end{figure}
Dose-response modeling and correlation analysis were performed for all four blast metrics and seven threshold levels to identify peak overpressure thresholds for physiologically significant blast events.  Figure~\ref{fig:DoseResponseThreshold} shows Spearman's correlations of cumulative dose levels for each session with the maximum physiology-based risk score (response), obtained by fusing the gait, balance, and blink feature modalities as described previously. Given that physiology may change over time, for example due to fatigue, we additionally trained a model using exposure duration as a feature rather than measured physiology, as indicated by the black reference line on the figure. 

Energy-based measures were relatively unaffected by removing the lowest peak-pressure events as expected, as these events contributed little to the accumulating total blast level. Dose-response correlations were also roughly the same as correlation with total exposure duration.
In contrast, level-based metrics produced substantially higher maximum (over thresholds) dose-response correlation than total exposure duration. Blast count improved, as expected, as the lowest-level events were removed, showing the highest maximum correlation overall at a threshold of 160 dB SPL. 
Importantly,  all acoustic measures showed a rapid reduction in dose-response correlation at a threshold level of 170 dB SPL, suggesting that capturing only high peak-pressure events (e.g., 1 PSI) may be inadequate to understand physiological response.

The best-performing blast measures showed significantly higher correlation with risk scores than blast duration alone. Further, we computed correlations between exposure time duration and the four blast outcome measures at the 160 dB threshold level, finding correlations ranging between 0.24 and 0.36, which is the same level of correlation we found between time duration and physiology-based risk scores. Together, these results support blast exposure as a direct predictor of physiology changes, rather than an indirect  correlation with some alternative predictor like fatigue which also increases with time duration.
\begin{table}[h!]
    \centering
    \begin{tabular}{l|c|c|c|c}
    Cumulative Blast Measure& Peak Pressure
(PSI)& Blast Count& Positive Impulse
(PSI-ms)& LZeq8hr
(dB)\\ \hline
    Peak Pressure& 1& -& -& -\\
    Blast Count& 0.94& 1& -& -\\
    Positive Impulse& 0.86& 0.80& 1& -\\
 LZeq8hr& 0.79& 0.71& 0.96&1\\
    \end{tabular}
    \caption{Spearman's correlation of per-session cumulative blast metrics (160 dBSPL event threshold). Positive Impulse/LZeq8hr and Peak Pressure/Count had the highest correlation coefficients }
    \label{tab:blast_metrics_spearman}
\end{table}
%Correlations above 0.21 and 0.27 are significant at the p=0.05 and p=0.01 level, respectively. 

\subsection{Wearable Sensor Ablation Study}

Fieldability is a primary consideration for a wearable blast monitoring system; while adding additional sensors may result in improved performance, incremental performance gains may come at the cost of battery life or end-user acceptance. The relative importance of each sensor in the LifeLens system was studied by comparing three sensor configurations: accelerometry-only (gait and balance features), accelerometry + single-channel EOG (+ blink features), and accelerometry + dual-channel EOG (+ saccade features). 
Table~\ref{tab:sensor_ablation} shows the results of these three sensor configurations for the two most effective blast exposure measures (Blast Count and Cumulative Peak Pressure) at two exposure threshold levels (155 dB and 160 dB). Shown are Spearman correlations of maximum per session model risk scores with actual exposure levels. Adding 1 EOG sensor improves the risk scores at the 160 dB threshold but not the 155 dB threshold. Adding an additional EOG sensor (which is only available in a minority of 36 sessions) to obtain saccade duration features does not improve prediction accuracy above that obtained with an accelerometer and a single EOG sensor. 
%Moreover, a benefit of the single EOG sensor is seen at the 160 dB threshold but not the 155 dB threshold.

\begin{table}
    \centering
    \begin{tabular}{l|c|c|c|c}
    Sensor  & Blast Cnt & Cum. PP & Blast Cnt & Cum. PP\\
    Modality & (155 dB) & (155 dB) & (160 dB) & (160 dB) \\ \hline
    Accel     & 0.52 & 0.51 & 0.53 & 0.48 \\
    1 EOG     & 0.18 & 0.14 & 0.35 & 0.21 \\
    2 EOG     & 0.20 & 0.18 & 0.36 & 0.20 \\
    Accel + 1 EOG & 0.49 & 0.51 & 0.60 & 0.52 \\
    Accel + 2 EOG & 0.49 & 0.49 & 0.58 & 0.48\\
    \end{tabular}
    \caption{Sensor ablation study. Spearman correlations, computed across 91 sessions, are shown between the model's maximum risk scores and the final blast measures for those sessions. Each row shows results from fusing a different number of sensor modalities.  }
    \label{tab:sensor_ablation}
\end{table}

\subsection{Direction of Correlations of Physiological Features Changes with Blast Exposure}

The above results show the magnitude of positive correlations between regression model outputs and blast exposure level, computed across 91 sessions. The direction of correlations of changes in individual physiological features with increasing exposure, computed over time within each session as well as across all sessions, is also of interest. The acceleration balance feature (acceleration path length) and blink duration are single channel, so it is straightforward to interpret the correlations of their changes with blast exposure over time.
The gait TDE features have high dimensionality (21-d eigenvalue vectors at each of 5 delay scales),  which makes it more difficult to interpret the correlations of their changes with blast exposure. The eigenvalues at each delay scale describe the shape of the 21-d TDE scatter distribution at that scale. Lower complexity dynamics are typically associated with a reduction in dimensionality, which is indicated by smaller eigenvalues at higher ranks (where eigenvalues are ranked from largest to smallest). We found the strongest correlations involved reductions in eigenvalue size correlating with blast exposure at the middle ranks (approximately ranks 6 through 14). To represent the change in complexity with a single value, we selected change in the eigenvalue at the middle of this range (rank 10), at the first delay scale.
The correlations in Table~\ref{tab:correlation_direction} indicate the same correlation sign for all four blast exposure measures, with the strongest correlation magnitudes for Blast Count and Cumulative Peak Pressure. The results show that blast exposure is associated with a lowering of gait complexity and a reduction in blink durations. Exposure is also associated with an increase in path length in the LM frames. Because these frames are selected based on a fixed acceleration magnitude variance threshold applied over a 5 s window, this finding suggests that blast exposure is associated with an increase in high time-frequency movements.

\begin{table}
    \centering
    \begin{tabular}{l|c|c|c|c|c}
    Feature  & \# data pts & Blast Cnt & Cum. PP & Total PI & LZeq\\ \hline
    Gait (TDE eig. rank 10)   & 1,977 & -0.22 & -0.23 & -0.20 & -0.15 \\
    Balance (LM path length) & 3,095 & 0.36 & 0.31 & 0.26 & 0.24 \\
    Blink duration & 34,129 & -0.16 & -0.14 & -0.10 & -0.06\\
    \end{tabular}
    \caption{Correlations are shown of single feature changes for gait, balance, and blinks, with the four blast exposure metrics at the 160 dB threshold. These are correlations from all time points across the 91 sessions. }
    \label{tab:correlation_direction}
\end{table}

\section{Discussion}

Our approach demonstrates that it is possible to develop acute dose/response models for blast exposure by combining accumulating exposure measures and changes in physiology.  This technology could be developed into a system to identify individual users who have a high cumulative exposure energy \emph{and} a physiological change as the most at risk.  We also address issues of individual susceptibility through individualized models.  
% , notification before injury occurrence, and interpretation of blast metrics for different weapon systems, this system would provide a major step forward from previous technologies.
% It is our view that on-body, objective acoustic measurements are best suited to account for these factors, but the question remains of what the best measures are for predicting brain injury, and at what levels and number of exposures injury is likely to occur.
% Historically, the scientific studies of blast exposure have focused on single events with the potential to cause harm through multiple mechanisms (i.e. direct vs. indirect blast effects). Meanwhile, both in training and combat scenarios, warfighters are exposed to sub-clinical blasts that have no obvious demonstrable physical effect but that seem to accumulate to cause brain health issues.

This work extends previous results that showed an ability to estimate blast exposure from EOG-derived features of eye blinks and saccades~\cite{rao2023changes} and accelerometry-derived gait features~\cite{williamson2022using}. Those results differed in being based on smaller data sets (31 sessions from 17 participants in \cite{rao2023changes} and 44 sessions from 17 participants in \cite{williamson2022using}). Additionally, in both previous studies, monotonic physiological changes across entire sessions were computed post hoc as a basis for regression models to estimate exposure level. 
The current work differs in five fundamental ways. First, a larger data set consisting of 91 sessions from 28 participants was included. 
Second, continuous-time tracking of physiological changes was used (Section~\ref{sec:feature-change}), which is sensitive to rapid physiological changes and can provide real-time alerts.
Third, the same physiological-change and regression modeling methods were applied to both EOG-based and accelerometer-based features, and the resulting models were fused.
Fourth, four different blast exposure metrics were evaluated and compared (Section~\ref{sec:dose-response}). Finally, continuous physiology was interpreted alongside cognitive assessments (ANAM). 

Brain health assessments like the ANAM, when conducted before and after weapons training, have the potential to measure the cognitive effects from brain injury. However, even in the case where a pre-/post-exposure test is sensitive to brain injury, a real-time monitoring approach can still localize  significant events (in time or dose) more accurately. We believe that continuous monitoring of the physiology and exposure, as presented in this study, can provide a more granular indication of the injury dose.  In a future fielded system, the ability to detect these changes in real time could prevent additional exposures after acute injury on an individualized basis, thus reducing the risk of further brain injury while maintaining high levels of training effectiveness.

Within this framework, we tracked physiology changes over time and with accumulating exposure, and identified a case study subject with significantly deteriorating ANAM scores over time that consistently showed rapidly deteriorating physiology with blast exposure over a span of multiple years, particularly in EOG-based features. The case-study subject's  anomalous changes in blink and balance physiology typically occurred several hours before the end of each day's training. Importantly, this analysis also provides a potential hazard threshold for that individual, who appeared to be unusually susceptible to low-level exposure. The case-study subject's consistent and anomalous changes in blink and balance physiology typically occurred several hours before the end of each day's training, supporting the need for tracking physiology during the training day. A delay in administering a differential assessment with comparable sensitivity would provide opportunity for additional exposure to accumulate, increasing the risk for further brain injury.  

When analyzing dose/response data from the entire cohort, we compared accumulating blast measures and thresholds for significant blast events for this group and environment by correlating blast measures with model predictions based on physiology. We found, using individualized machine learning models trained using a "leave-one-out" approach, that accelerometry scores provided good correlation with cumulative blast exposure measures, with a small but noticeable benefit from fusing EOG features. Despite this relatively small impact, we believe EOG may still be impactful for early warning of injury based on the case-study findings, though additional insights may be gained from accelerometry alone. Likewise, tracking gaze or vestibulo-ocular reflex (VOR) using a higher-fidelity sensor could prove impactful.

Our dose-response correlation analysis fusing per-session blink and balance scores further identified candidate blast features and thresholds to identify significant low-level blast events. In particular, a count of blast events with a measured peak pressure above 160 dB SPL showed the highest group dose-response correlation overall. This "blast count" measure, however, was the most sensitive feature to threshold level, and therefore an accumulating peak pressure could provide a more robust dose metric in general while maintaining good correlations in this group, as it is less sensitive to low-pressure events that may have little impact on physiology. Still, in limited cases, a lower fidelity sensor, if properly designed and calibrated, may prove useful in tracking blast exposure by counting events. In all cases, removing blast events greater than 170 dB SPL as measured by the MNOISE resulted in rapid reduction in dose-response correlation, indicating the importance choosing a device that can measure and record these low-level events.

It would also be of interest to assess the ability of the physiological model to predict same-day reductions in ANAM scores, relative to predictions based on measured blast exposure levels. We analyzed 23 sessions with physiological recordings occurring on the same day that same-day changes in ANAM scores were available. The correlations of measured blast exposure with these same-day changes in ANAM were statistically insignificant for all of the blast exposure measures and thresholds. 
 
In summary, our study demonstrates a dose-response relationship based on real-time continuous monitoring of blast exposure and acute physiological response from wearable devices, and demonstrates the practical value of continuous monitoring relative to pre/post exposure assessments.  We additionally found a threshold (0.25 PSI) for significant blast events producing the highest dose/response correlation that is considerably lower than the minimum peak pressure threshold on some commercial blast gauges. Although this study demonstrates a candidate physiological wearable approach, we recognize that these results represent a limited data sample and may apply only to a specialized cohort; further development is needed to develop and validate this approach as a decision aid in other overpressure environments.

% Therefore, improved blast measures are needed that combine exposure levels for multiple blast events in some way. Likewise, individual susceptibility to blasts and the impact of weapon systems are likely important factors; different weapon systems may pose unique risks for the same level or number of exposures, and two individuals exposed to similar blast events may respond differently, both acutely and chronically. In the context of this uncertainty a difficult trade-off must be undertaken by commanders; too many exposures risks long-term brain health, while too few could jeopardize the effectiveness of training.

% \subsection{Scientific Relevance}

\subsection{Translation to Combat Environments }

Wearable physiology monitoring brings unique benefits relative to assessments in training environments; for example, the potential to capture multiple measures of physiology simultaneously and to localize significant physiological responses in time (or exposure level).  These benefits become even more apparent in austere combat settings where pre- / post-exposure physiological assessments can be difficult or impossible to administer, for example, because blast exposure periods may be unpredictable. 
We suggest that the most practical approach is to develop these technologies in training settings alongside assessments closely linked to brain injury. These findings could then directly support wearable monitoring to identify and prevent blast injury in combat environments. 

\subsection{Limitations and Future Work}

Our work indicates that pairing continuous monitoring with periodic assessments like ANAM has the potential to provide range commanders with evidence to prevent accumulating brain injury during occupational training in the near-term. However, we acknowledge several limitations of physiology monitoring in general relative to blast exposure monitoring that we believe assessments like ANAM may address. First, physiology change is non-specific to brain injury generally. Second, a clear link between acute changes in physiology and acute brain injury leading to long-term health effects has not yet been established. Reaction times, for example,  increasing significantly over several weeks or months could be related directly to long-term brain health effects, whereas acute physiology changes may not. Still, continuous physiology scores can localize potential injuries to the day or even minutes and are more closely related to the accumulated exposure level at that time, making them an important tool for identifying safe exposure limits.
Based on these observations and our results, we conclude that an early warning system combining real-time monitoring (of exposure and physiology) with neuro-cognitive assessments could leverage the strengths of each approach, providing better evidence for occupational health decisions than either individually.  

%As mentioned, one important limitation of our approach is that to date, relationships between blast overpressure exposure, acute physiological response, and long-term brain health deficits are unknown or not well understood. While multiple on-going research efforts attempt to address these basic science questions

%To test our hypothesis, we compared individualized dose/response model predictions with simple reaction reaction time scores from ANAM tests for this cohort. 

We also  interpret our results to be limited in terms of multiple experimental factors, including type of weapon/exposure, environmental conditions, and cohort (i.e., instructors are highly specialized), and range of the exposures that are encountered. As a result,  particular dose/response findings may not generalize to other training environments. Particularly, although we found the highest dose/response correlation using blast count (Table \ref{tab:correlation_direction}), we would anticipate in a larger more diverse dataset an energy based metric or other cumulative metric could result in higher correlations \cite{mcevoy2024cumulative}.

Nevertheless, these results motivate further work including wearable monitoring technology to uncover safe exposure thresholds in training environments and to identify susceptible individuals who may experience injury at levels that are safe for other people. 

\section*{Conclusion}

Wearable technology can be used to simultaneously and passively monitor both blast exposure and resulting physiological changes. We found that a model relating exposure (dose) to physiological response revealed that blast events with peak pressure levels as low as 0.25 PSI are related to physiological changes and could contribute to blast injury.  We also identified an individual subject with deteriorating reaction-time scores that consistently showed a rapid and anomalous change in physiology-based risk scores after as few as 12 low-level blast events in a given session,  supporting our hypotheses that acute brain injury leading to long-term brain health effects results in measurable changes in physiology and that susceptibility to low-level blast injury depends on as-yet unknown individual factors.  Overall, our results suggest that the wearable approach to blast monitoring can lead to faster intervention, and a better understanding of the most appropriate metrics for exposure dose (combining number and intensity of exposures), as well as (possibly individualized) safe levels are needed to protect the brain health of warfighters in these environments. 

% \section*{Data Availability}
% The data are available from the corresponding author on reasonable request.

% \section*{Code Availability}
% The code are available from the corresponding author on reasonable request.

%\clearpage
\section*{Acknowledgements}
The authors would like to acknowledge MAJ Kurtis Gruters, Dr. Katherine Spradley, Mr. Sedrick Thomas, Dr. Luanne Kent, MSG David Curtis, and SFC Aaron Anderson from USASOC as well as Dr. Andrea Vincent, Vista LifeSciences, Creare, and LifeLens for hardware and data support. 

\section*{Funding}
This material is based upon work supported by the United States Air Force under Air Force Contract No. FA8702-15-D-0001 and by the United States Army Medical Materiel Development Activity (USAMMDA).

\section*{Distribution}

DISTRIBUTION STATEMENT A. Approved for public release. Distribution is unlimited. This material is based upon work supported by the Department of the Army under Air Force Contract No. FA8702-15-D-0001. Any opinions, findings, conclusions or recommendations expressed in this material are those of the author(s) and do not necessarily reflect the views of the Department of the Army.

% DISTRIBUTION STATEMENT C. Distribution authorized to U.S. Government Agencies and their contractors; Test \& Evaluation; August 26th, 2016. Other requests for this document shall be referred to U.S. Army Medical Materiel Development Activity, 1430 Veterans Drive, Ft Detrick, MD 21702. WARNING: This document may contain technical data whose export is restricted by the Arms Export Control Act (AECA) or the Export Control Reform Act of 2018 (ECRA). Transfer of this data by any means to a non-US person who is not eligible to obtain export-controlled data is prohibited. By accepting this data, the consignee agrees to honor the requirements of the AECA and ECRA. DESTRUCTION NOTICE: For unclassified, limited distribution documents, destroy by any method that will prevent disclosure of the contents or reconstruction of the document.This material is based upon work supported by the Dept of the Army under Air Force Contract No. FA8702-15-D-0001. Any opinions, findings, conclusions or recommendations expressed in this material are those of the author(s) and do not necessarily reflect the views of the Dept of the Army.© 2025 Massachusetts Institute of Technology. Delivered to the U.S. Government with Unlimited Rights, as defined in DFARS Part 252.227-7013 or 7014 (Feb 2014). Notwithstanding any copyright notice, U.S. Government rights in this work are defined by DFARS 252.227-7013 or DFARS 252.227-7014 as detailed above. Use of this work other than as specifically authorized by the U.S. Government may violate any copyrights that exist in this work. 

\section*{Competing Interests}
The authors declare that there are no competing interests.

\section*{Human Subjects}
The USASOC Exemption Determination Officer reviewed and determined the collection of data through this program evaluation to be Not Research IAW (32 CFR 219). Data analysis was found not to meet the federal definition of human subjects research by MIT COUHES (45 CFR 46).

\section*{Disclaimer}
The views expressed in this article are those of the authors and do not necessarily reflect the views of US Army Special Operations Command, the Department of the Army, or the Department of Defense. 
%\nocite{*} % this makes all the references appear even if you havent yet cited them, to get started

%\bibliographystyle{naturemag}
%\bibliography{references}
\printbibliography

% \renewcommand{\cftfigpresnum}{Figure }
% \renewcommand{\cftfignumwidth}{6em}
% \renewcommand{\cftfigaftersnum}{:}

%  \addtolength{\cfttabnumwidth}{3em}
%  \renewcommand{\cfttabpresnum}{\tablename\ }
% \renewcommand{\cfttabaftersnum}{:} 
% %To remove dots from the list of figures and list of tables
 % \makeatletter \renewcommand{\@dotsep}{10000} \makeatother
\clearpage
\pagenumbering{gobble}
% \listoffigures
\end{document}